%% file: dpcArxiv.tex
\documentclass{article}
\usepackage{epsf,psfrag, subfigure,epsfig, times,amsmath,amssymb,color}

\usepackage{ew2010}                       

\title{Secret key agreement on wiretap channels with transmitter side information}

\name{Ashish Khisti}

\address{ University of Toronto \\
Toronto, ON, Canada \\
phone: + (1) 416-978-7215, fax: + (1) 416-978-4425, email: akhisti@comm.utoronto.ca \\
web: http://www.comm.utoronto.ca/~akhisti }

\input{gww_chars}

\input{gww_defs}

\input{gww_rvs}

\def\blfootnote{\xdef\@thefnmark{}\@footnotetext}

\renewcommand{\rvk}{\mathsf{\kappa}}

\newcommand{\rvsr}{\mathsf{s}_{r}}

\begin{document}

\maketitle

\begin{abstract}
Secret-key agreement protocols over wiretap channels controlled by
a state parameter are studied. The entire state sequence is known
(non-causally) to the sender but not to the receiver and the
eavesdropper. Upper and lower bounds on the secret-key capacity are
established both with and without public discussion. The proposed coding scheme involves constructing a codebook to  create common reconstruction of the state sequence
at the sender and the receiver and another secret-key codebook constructed by random binning. 
For the special case of Gaussian channels, with no public discussion, --- the \emph{secret-key generation with dirty paper} problem, the gap between our bounds is at-most 1/2~bit and the bounds coincide in the high signal-to-noise ratio and high interference-to-noise ratio regimes. 
In the presence of public discussion our bounds coincide, yielding the capacity, when then the channels of
the receiver and the eavesdropper satisfy an independent noise
condition.
\end{abstract}

\section{Introduction}
Many applications in cryptography require that the legitimate
terminals have shared secret keys, not available to unauthorized
parties. Information theoretic security encompasses the study of
source and channel coding techniques to generate secret-keys between
legitimate terminals. In the channel coding literature, an early
work in this area is the wiretap channel
model~\cite{wyner:75Wiretap}. It consists of three terminals — one
sender, one receiver and one eavesdropper. The sender communicates
to the receiver and the eavesdropper over a discrete-memoryless
broadcast channel. A notion of equivocation-rate — the normalized
conditional entropy of the transmitted message given the observation
at the eavesdropper, is introduced, and the tradeoff between
information rate and equivocation rate is studied. Perfect secrecy
capacity, defined as the maximum information rate under the
constraint that the equivocation rate approaches the information
rate asymptotically in the block length is of particular interest.
Information transmitted at this rate can be naturally used as a
shared secret-key between the sender and the receiver. In the source
coding setup~\cite{maurer:93,ahlswedeCsiszar:93} the two terminals
observe correlated source sequences and use a public discussion
channel for communication. Any information sent over this channel is
available to an eavesdropper. The terminals generate a common
secret-key that is concealed from the eavesdropper in the same sense
as the wiretap channel — the equivocation

In the present paper we consider a secret-key agreement problem when
the sender and the receiver communicate over a channel controlled by
a state parameter. The state parameter is known to the sender but
not to the receiver or the eavesdropper.   The problem of transmitting information
on such channels, without secrecy constraints, is studied in~\cite{gelfandPinsker:80}. 
A random binning strategy is proposed and shown to achieve the capacity. Costa~\cite{costa:83} studies the problem of
communicating over an additive noise Gaussian channel with an additive interference sequence known to the transmitter and establishes that there is no loss in capacity if the interference sequence is known only to the transmitter.
In the present work, we study the problem of generating a common secret key between the sender and the receiver over such channels. Our proposed coding scheme is not based on the Gel'fand-Pinsker binning technique for sending an information message
over such channels. Instead our codebook is designed to create a common reconstruction sequence at the sender and the receiver
and distilling a secret-key based on this common sequence.

In related works, the problem of secret-message transmission over
wiretap channels controlled by a state parameter is studied
in~\cite{mitrapant:06,LiuChen:07}. In these works an achievable
coding scheme is proposed that combines Gel'fand Pinsker coding and
coding for the wiretap channel. As discussed earlier, our coding scheme is 
based on a different approach and in general yields higher achievable rates.
A related  problem of common reconstruction of state sequences has been studied recently in
\cite{steinberg:08, steinberg:09}. The problem of secret-key agreement with symmetric channel state information
at the sender and the legitimate receiver has been studied in~\cite{khistiDigWornell:09}. However the coding scheme involved
is based on the fact that the terminals have knowledge of the common state sequence to begin with. After this paper
was submitted, we learnt about a recent work~\cite{prabhakaran:09} where the problem of communicating over
over channels with non-causal CSI is used as a building block for characterizing the tradeoff between secret-key
and secret-message transmission.

\section{Problem Setup}
\begin{figure*}
\centering \psfrag{sr}{$\rvs$}\psfrag{k}{$\rvk$}
\includegraphics[scale=0.5]{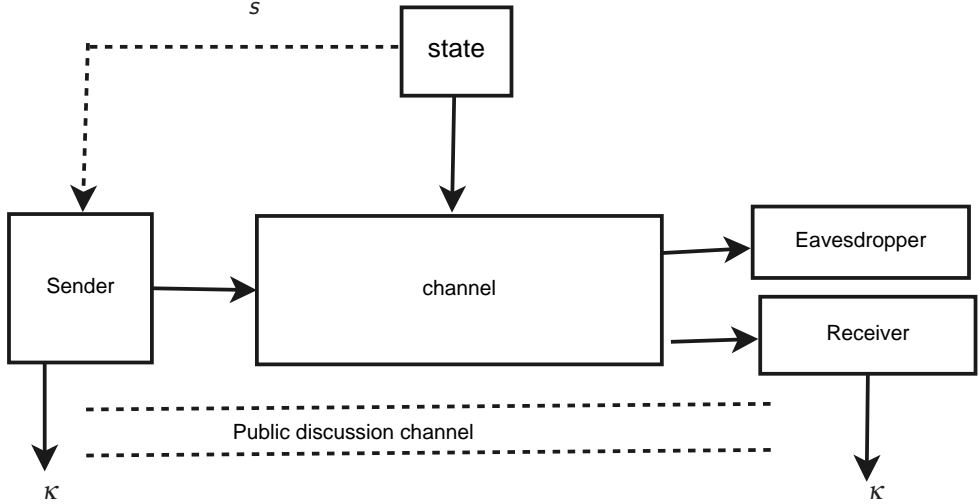}
\caption{Wiretap channel controlled by a state parameter. The
channel transition probability $p_{y_r,y_e|x,s}()$ is controlled by
a state parameter $\rvs$. The entire source sequence $\rvs^n$ is
known to the sender but not to the receiver or the eavesdropper. The
sender and receiver generate a secret key $\rvk$ at the end of the
transmission.}\label{fig:wiretap}
\end{figure*}

As Fig.~\ref{fig:wiretap} illustrates, the channel model has three
terminals
--- a sender, a receiver and an eavesdropper.
 The sender communicates with the other two terminals over a
discrete-memoryless-channel with transition probability $p_{\rvyr,
\rvye|\rvx,\rvs}(\cdot)$ where $\rvx$ denotes the channel input
symbol, whereas $\rvyr$ and $\rvye$ denote the channel output
symbols at the receiver and the eavesdropper respectively. The
symbol $\rvs$ denotes a state variable that controls the channel
transition probability. We assume that it is sampled i.i.d.\ from a
distribution $p_{\rvs}$ in each channel use. Further, the entire
sequence $\rvs^n$ is known to the sender before the communication
begins.

In defining the secret-key capacity we separately consider the cases
when a public discussion channel is and is not present.
\subsection{No Public Discussion} \label{subsec:noDisc} A
length $n$ encoder is defined as follows. The sender samples a
random variables $\rvu$ from the conditional distribution
$p_{\rvu|\rvs^n}(\cdot|s_r^n)$. The encoding function produces a
channel input sequence $\rvx^n = f_n(\rvu, \rvs^n)$ and transmits it
over $n$ uses of the channel. At time $i$ the symbol $\rvx_i$ is
transmitted and the legitimate receiver and the eavesdropper observe
output symbols $\rvy_{ri}$ and $\rvy_{ei}$ respectively, sampled
from the conditional distribution
$p_{\rvyr,\rvye|\rvx,\rvs}(\cdot)$. The sender and receiver compute
secret keys $\rvk = g_n(\rvu,\rvs^n)$ and $\rvl = h_n(\rvyr^n)$. A
rate $R$ is achievable if there exists a sequence of encoding
functions such that for some sequence $\eps_n$ that vanishes as
$n\rightarrow\infty$, we have that $\Pr(\rvk\neq\rvl) \le \eps_n$
and $\frac{1}{n}H(\rvk) \ge R - \eps_n$ and \begin{equation}
\label{eq:secNoInterac} \frac{1}{n}I(\rvk;\rvye^n) \le \eps_n.
\end{equation} The largest achievable rate is the secret-key capacity.

\subsection{Presence of Public Discussion}
\label{subsec:Disc} When a public discussion channel is present, the
described protocol follows closely the interactive communication
protocol  in~\cite{ahlswedeCsiszar:93}. The sender transmits symbols
$\rvx_1, \ldots, \rvx_n$ at times $0 < i_1 < i_2 < \ldots < i_n$
over the wiretap channel. At these times the receiver and the
eavesdropper observe symbols $\rvy_{r1}, \ldots , \rvy_{rn}$ and
$\rvy_{e1}, \ldots , \rvy_{en}$ respectively. In the remaining times
the sender and receiver exchange messages
 $\psi_t$ and  $\phi_t$ where $1 \le t \le k$. For convenience we let $i_{n+1} = k +1$. The eavesdropper observes
both $\psi_t$ and  $\phi_t$.

More specifically the sender and receiver sample random variables
$\rvu$ and $\rvv$ from conditional distributions
$p_{\rvu|\rvs^n}(\cdot|s_r^n)$ and $p_{\rvv}(\cdot)$ and observe
that $\rvv$ is independent of $(\rvu,\rvs^n)$.
\begin{itemize}
\item  At times $0 < t < i_1$, the sender generates $\phi_t = \Phi_t(\rvu, \rvs^n,  \psi^{t-1})$ and the receiver
generates  $\psi_t =    \Psi_t(\rvv,  \phi^{t-1})$. These messages
are exchanged over the public channel.
\item At times $i_j$ , $1 \le j \le n$, the sender generates $\rvx_j = X_j(\rvu, \rvs^n,  \psi^{i_j-1})$ and sends it over
the channel. The receiver and eavesdropper observe $\rvy_{r,j}$ ad
$\rvy_{e,j}$ respectively. For these times we set $\psi_{i_j} =
\phi_{i_j} = 0$.
\item For times $i_j < t < i_{j+1}$, where $1 \le j \le n$, the sender and receiver compute $\phi_t =
\Phi_t(\rvu, \rvs^n,  \psi^{t-1})$ and  $\psi_t =  \Psi_t(\rvv ,
\rvyr^j , \phi^{t-1})$ respectively and exchange them over the
public channel.
\item At time $k + 1$, the sender and receiver compute $\rvk = g_n(\rvu, \rvs^n,\psi^k)$ and the receiver
computes $\rvl = h_n(\rvv , \rvyr^n, \phi^k)$.
\end{itemize}

We require that for some sequence $\eps_n$ that vanishes as $n
\rightarrow \infty$, $Pr(\rvk \neq \rvl ) \le \eps_n$ and
\begin{equation}\frac{1}{n}I(\rvk;\rvye^n,\psi^k,\phi^k) \le \eps_n.\label{eq:secInterac}\end{equation}

The secret-key rate is defined as $\frac{1}{n}H(\rvk)$ and the
largest achievable secret-key rate is the capacity.

\section{Main Results}

Our main results are upper and lower bounds on the secret-key
capacity, which coincide in some special cases. We again consider
the cases of no public discussion and public discussion separately.

\subsection{No Public Discussion}
We first provide an  achievable rate (lower bound) on the secret-key
capacity.
\begin{thm}
An achievable secret-key rate without public discussion is
\begin{equation}
\label{eq:lowerBoundNoDisc} R^- = \max_{p_\rvu, p_{\rvx|\rvs,\rvu}}
I(\rvu;\rvyr) - I(\rvu;\rvye),
\end{equation}
where the maximization is over all auxiliary random variables $\rvu$
that satisfy the Markov condition $\rvu \rightarrow (\rvx, \rvs)
\rightarrow (\rvyr,\rvye)$ and furthermore satisfy the constraint
that
\begin{equation}
I(\rvu;\rvyr)-I(\rvu;\rvs) \ge 0. \label{eq:lowerBoundNoDiscCons}
\end{equation}\label{thm:lbNoDisc}
\end{thm}

The intuition behind the coding scheme is as follows. Upon observing
$\rvs^n$, the sender communicates the best possible reproduction
$\rvu^n$ of the state sequence to the receiver. Now both the sender
and the receiver observe a common sequence $\rvu^n$. The set of all
codewords $\rvu^n$ is binned into $2^{nR^-}$ bins and the bin-index
is declared to be the secret key.

We note that the lower bound can be easily extended to the case of
two-sided CSI. If the receiver observes another state sequence
$\rvsr,$ correlated with $\rvs$ according to a joint distribution
$p_{\rvs,\rvsr}(\cdot,\cdot)$ then the achievable rate
expression~\eqref{eq:lowerBoundNoDisc} holds provided that we
augment the received symbol by $(\rvyr,\rvsr)$.

Finally for the case of symmetric CSI i.e., when $\rvsr = \rvs$, the
constraint~\eqref{eq:lowerBoundNoDiscCons} is redundant as clearly
$I(\rvu;\rvyr,\rvs)-I(\rvu;\rvs) \ge 0$ holds. Furthermore the
resulting achievable rate, $$R^- = \max_{p_\rvu, p_{\rvx|\rvs,\rvu}}
I(\rvu;\rvyr,\rvs) - I(\rvu;\rvye)$$ is indeed the secret-key
capacity as established in our earlier
work~\cite{khistiDigWornell:09}.

Finally we note that the problem of secret-key agreement is
different from the secret-message transmission problem considered
in~\cite{mitrapant:06,chenVinck:06,LiuChen:07}. This is because the
secret-key can be an arbitrary function of the state sequence (known
only to the transmitter) whereas the secret-message needs to be
independent function of the state sequence. For comparison, the best
known lower bound on the secret-message transmission problem is
stated below.
\begin{prop}\cite{mitrapant:06,chenVinck:06,LiuChen:07}
An achievable secret message rate for wiretap channel with
non-causal transmiter CSI is
\begin{equation}
R \le \max_{p_{\rvu},p_{\rvx|\rvu,\rvs}}I(\rvu;\rvyr) -
\max\left(I(\rvu;\rvs), I(\rvu;\rvye)\right).\label{eq:secMsgRate}
\end{equation}
\end{prop}

We note that whenever the maximizing $\rvu$ satisfies,
$I(\rvu;\rvye)
> I(\rvu;\rvs)
> I(\rvu;\rvye)$, the secret-key rate~\eqref{eq:lowerBoundNoDisc} is
strictly better than the secret-message rate~\eqref{eq:secMsgRate}.

The following theorem develops an upper bound on secret-key capacity
that is amenable to numerical computation.

\begin{thm}
The secret-key capacity in absence of public discussion is upper
bounded by $C \le R^+$, where
\begin{equation}
\label{eq:upperBoundNoDisc} R^+ = \min_{p_{\rvyr,\rvye|\rvx} \in
\cP}\max_{p_{\rvx|\rvs}} I(\rvx,\rvs;\rvyr|\rvye),
\end{equation}
where $\cP$ denotes all the joint distributions
$p^\star_{\rvyr,\rvye|\rvx,\rvs}$ that have the same marginal
distribution as the original channel. \label{thm:ubNoDisc}
\end{thm}

The intuition behind the upper bound is as follows. We create a
degraded channel by revealing the output of the eavesdropper to the
legitimate receiver. We further assume a channel with two inputs
$(\rvx^n,\rvs^n)$ i.e., the state sequence $\rvs^n$ is not
arbitrary, but rather a part of the input codeword with distribution
$p_{\rvsr}$. The secrecy capacity of the resulting wiretap channel
is then given by $I(\rvx,\rvs;\rvyr|\rvye)$.

Our proposed upper and lower bounds coincide, yielding capacity in
some special cases. We present one such case in
section~\ref{subsec:Gaussian}.

\subsection{With Public Discussion}
In this section we provide lower and upper bounds on the secret-key
capacity with public discussion. We first provide a lower bound
below.

\begin{thm}
An achievable secret-key rate with public discussion is:
\begin{equation}
R^-_\mrm{disc} = \max\left( \max_{p_{\rvx|\rvs}} I(\rvx,\rvs;\rvyr)
- I(\rvye;\rvyr), R^- \right)
\end{equation}
where $R^-$ is the lower bound attained without public discussion in
Theorem~\ref{thm:lbNoDisc} \label{thm:lbDisc}
\end{thm}

The achievability scheme involves a natural modification of Maurer's
coding scheme~\cite{ahlswedeCsiszar:93,maurer:93} to incorporate the
presence of the state parameter and involves a single round of
discussion. In particular, the sender generate a sequence $\rvx^n$
according to the conditional distribution $p_{\rvx|\rvs}(x|s)$ and
transmits over $n$ channel uses. At the end of the transmission, the
receiver sends the bin index of $\rvyr^n$, so that the sender can
recover this sequence given $(\rvx^n,\rvsr^n)$.

Next we provide an upper bound on the secret-key capacity under
public discussion.

\begin{thm}
An upper bound on the secret-key capacity is
\begin{equation}
R^+ = \max_{p_{\rvx|\rvs}} I(\rvx,\rvs;\rvyr|\rvye)
\label{eq:TxCSIUB}.
\end{equation}\label{thm:ubDisc}
\end{thm}
We note that the upper bound expression~\eqref{eq:TxCSIUB} is
similar to the upper bound expression in~\eqref{eq:upperBoundNoDisc}
except that we cannot minimize over the joint-probability
distribution in~\eqref{eq:TxCSIUB}. This is because the public
discussion channel  provides a mechanism for feedback and hence the
capacity does depend on the joint distribution (not just the
marginal distributions). The proof for the upper bound expression in
Theorem~\ref{thm:ubDisc} also significantly more elaborate as it
accounts for public discussion.

We note that if the channel additionally satisfies $\rvyr
\rightarrow (\rvx,\rvs) \rightarrow \rvye$ then the upper and lower
bounds in Theorem~\ref{thm:lbDisc} and~\ref{thm:ubNoDisc} coincide.
In particular if $p_{\rvx|\rvs}$ is the maximizing distribution
in~\eqref{eq:TxCSIUB}, we have that
\begin{align*}
R^-_\mrm{disc} &\ge I(\rvx,\rvs;\rvyr) - I(\rvye;\rvyr) \\
&= I(\rvye,\rvx,\rvs;\rvyr) - I(\rvye;\rvyr) -
I(\rvyr;\rvye|\rvx,\rvs)\\
&=I(\rvye,\rvx,\rvs;\rvyr) - I(\rvye;\rvyr)\\
&= I(\rvx,\rvs;\rvyr|\rvye) = R^+_\mrm{disc}.
\end{align*} Since $R^-_\mrm{disc} \le R^+_\mrm{disc}$, it follows
that the two expressions must be equal. This is summarized in the
result below.
\begin{thm}
The secret-key capacity with public discussion for a DMC channel
that satisfies $\rvyr \rightarrow (\rvx,\rvs) \rightarrow \rvye$ is
given by
\begin{equation}
C_\mrm{disc}=\max_{p_{\rvx|\rvs}}
I(\rvx,\rvs;\rvyr|\rvye)\label{eq:capDisc}.
\end{equation}\label{thm:capdisc}
\end{thm}

\subsection{Gaussian Case}
\label{subsec:Gaussian} We now study the Gaussian special case under
an average power constraint. The channel to the legitimate receiver
and the eavesdropper is expressed as:
\begin{equation}\begin{aligned}
\rvyr &= \rvx + \rvs + \rvzr\\
\rvye &= \rvx + \rvs + \rvze,
\end{aligned}\label{eq:GaussianModel}\end{equation} where
$\rvzr \sim \cN(0,1)$ and $\rvze \sim \cN(0,1+\Delta)$ denote the
additive white Gaussian nose and are assumed to be sampled
independently. The state parameter $\rvs \sim \cN(0,Q)$ is also
sampled i.i.d.\ at each time instance and is independent of both
$\rvzr$ and $\rvze$. Furthermore, the channel input satisfies an
average power constraint $E[\rvx^2] \le P$. As the title indicates,
we call this setup, \it{secret sharing with dirty paper}.

Thus the parameter $P$ denotes the signal-to-noise ratio, the
parameter $Q$ denotes the interference-to-noise-ratio, whereas
$\Delta$ denotes the degradation level of the eavesdropper. We now
provide lower and upper bounds on the secret-key capacity with and
without public discussion. For simplicity in exposition we limit our analysis to the case when $P \ge 1$.

\begin{figure*}[!htb]
\begin{center}
\includegraphics[scale=0.5]{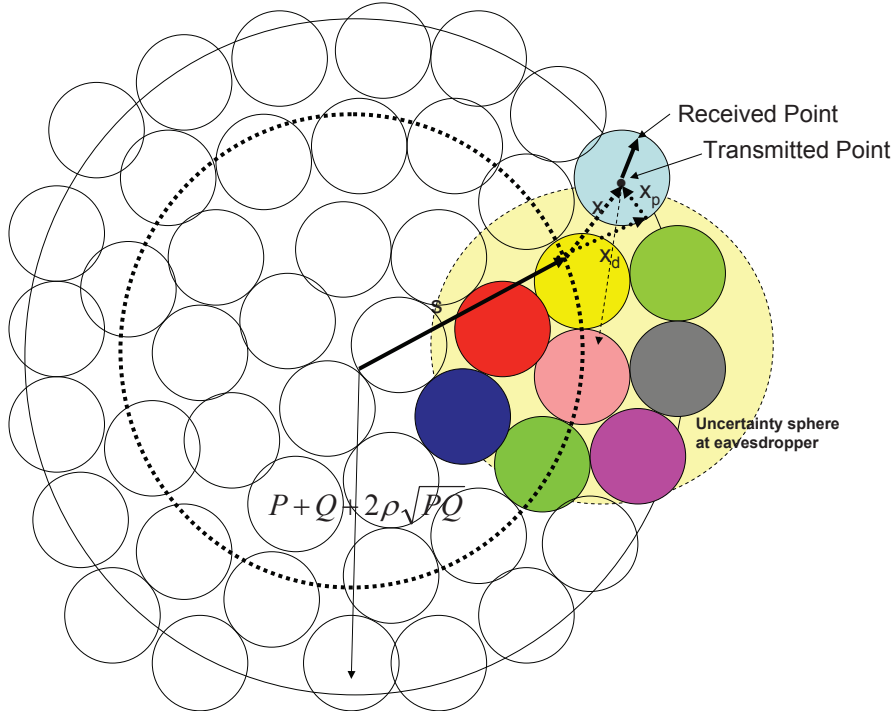}
\end{center}\label{fig:dpc}\caption{Secret-key agreement codebook for the dirty paper channel. The transmitter signal $\rvx^n$ is selected so that $\rvu^n = \rvx^n+\rvs^n$ is a sequence in the random codebook. The legitimate receiver can decode $\rvu^n$ (with high probability) and map it to the secret-key. The eavesdropper's noise-uncertainity sphere includes are possible key values.  Note that unlike the traditional dirty-paper code, the transmiter signal $\rvx^n$ has a component along $\rvs^n$. The achievable rate, does depend on the interference power and hence it is beneficial to amplify it using part of the transmit power. Also note that unlike a dirty-paper code we do not  scale down $\rvs^n$ before quantizing but use $\al =1$.}\end{figure*}

\begin{prop}
Assuming that $P \ge 1$, a lower bound on the secret-key agreement capacity is
capacity is given by,
\begin{multline}
R^- =  \frac{1}{2}\log\left(1 + \frac{\Delta(P+Q+2\rho \sqrt{PQ})}{P+Q+1+\Delta + 2\rho \sqrt{PQ}}\right),
\label{eq:lbGaussNoDisc}\end{multline}
where $\rho < 1$ is the largest value that satisfies
\begin{equation}
P(1-\rho^2) \ge 1 - \frac{1}{P+Q+1}.
\end{equation}
\label{prop:lbGaussNoDisc}
\end{prop}
\begin{prop}
In absence of public discussion, an upper bound on the secret-key
capacity is given by,
\begin{equation}
R^+ = \frac{1}{2}\log\left(1 +
\frac{\Delta(P+Q + 2\sqrt{PQ})}{P+Q+1+\Delta + 2\sqrt{PQ}}\right)\label{eq:ubGaussNoDisc}
\end{equation}\label{prop:ubGaussNoDisc}
\end{prop}
It can be readily verified that the upper and lower bounds coincide in several asymptotic regimes.
\begin{prop}
The upper and lower bounds on secret-capacity without public discussion satisfying the following
\begin{align}
&\forall P\ge 0,~R_+ - R_- \le \frac{1}{2} \label{eq:univGap} \\
&\lim_{P\rightarrow\infty} R_+ - R_- = 0 \label{eq:highINR}\\
&\lim_{Q\rightarrow\infty} R_+ - R_- = 0 \label{eq:lowINR}
\end{align}
\end{prop}

\begin{prop}
\label{prop:DPCwDC}
In the presence of public discussion, the secret-key capacity is given by the following expression,
\begin{equation}
R^+ = \frac{1}{2}\log\left(1 + \frac{(1+\Delta)(P+Q+2\sqrt{PQ})}{P+Q+1+\Delta + 2\sqrt{PQ}}\right)
\label{eq:DPCwDC}\end{equation}
\end{prop}

\begin{figure*}
\begin{minipage}[b]{0.5\linewidth}
\centering
\includegraphics[width=10cm]{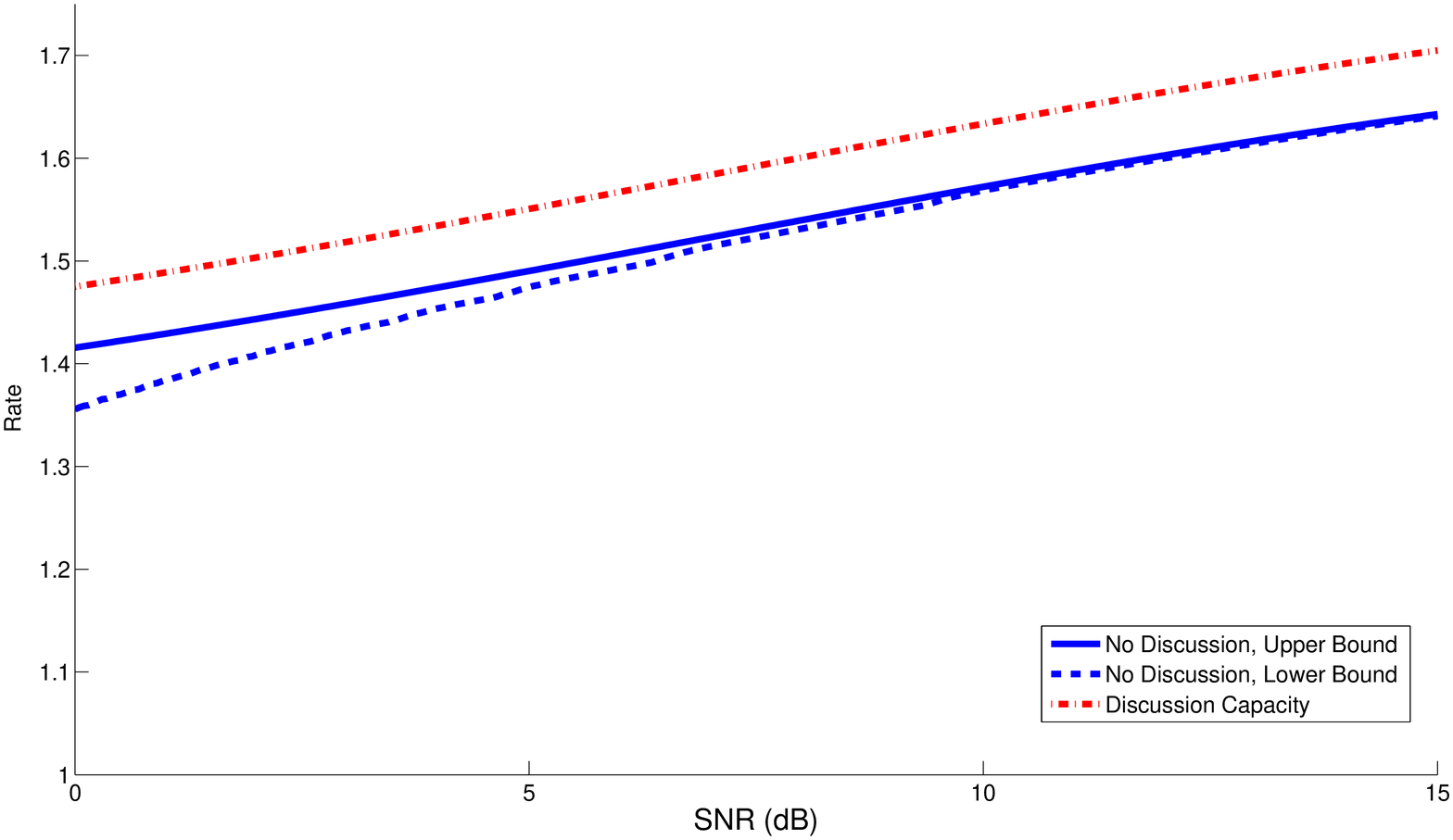}
\end{minipage}
\hspace{0.5cm} 
\begin{minipage}[b]{0.5\linewidth}
\centering
\includegraphics[width=10cm]{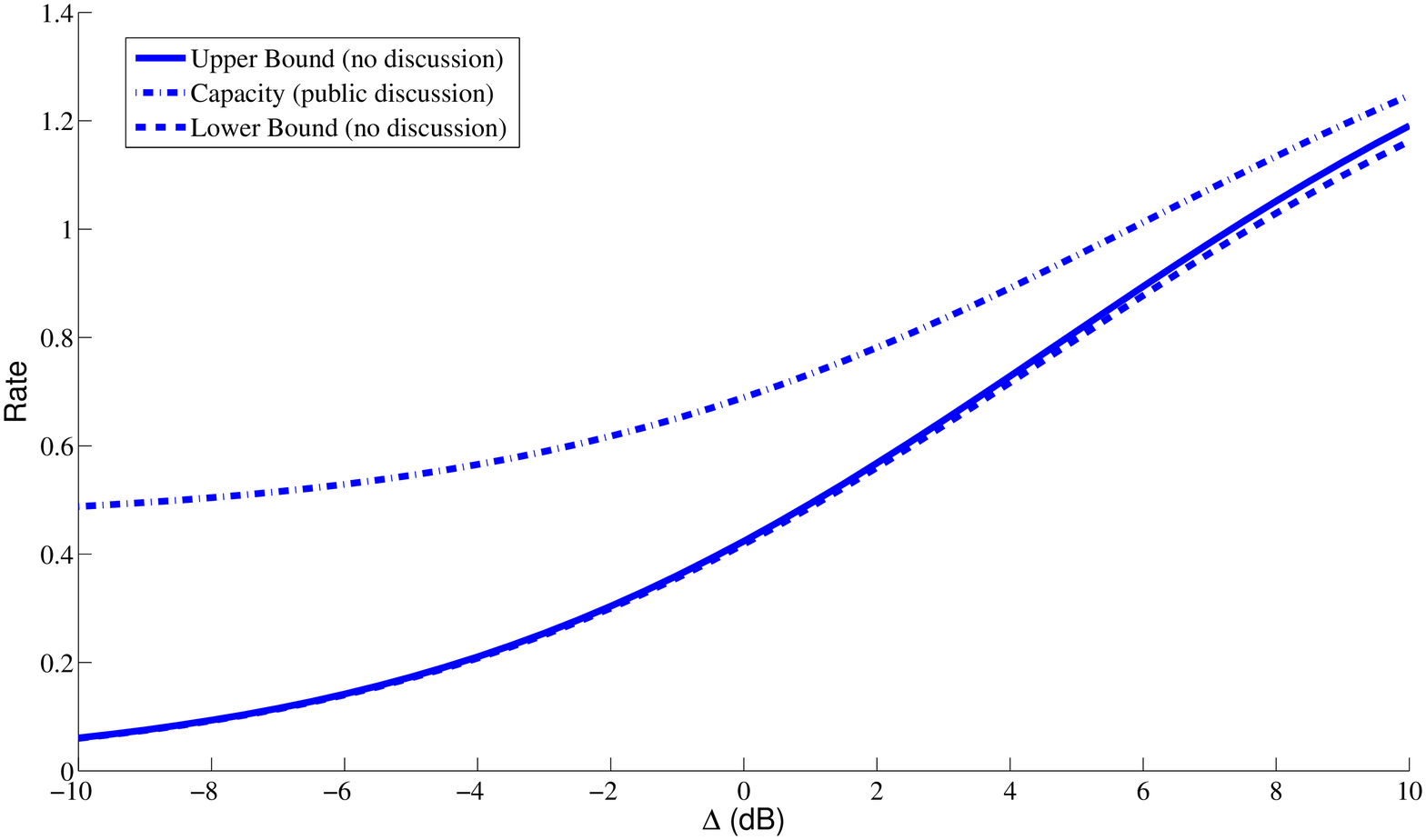}
\end{minipage}
\caption{Bounds on the capacity of the ``secret-sharing with dirty paper" channel
with and without public discussion. In the left figure, we
plot the capacity as a function of SNR (dB) when $Q = 10$ and
$\Delta = 10$. The upper-most curve is the capacity with
public-discussion whereas the other two curves denote the upper and
lower bounds without discussion. In
the right figure we plot the capacity with 
public discussion as a function of $\Delta$ (in dB) when $P=10$ dB
and $Q = 10$ dB as well as the upper and lower bounds without public discussion. }
\end{figure*}

\section{Without Public Discussion}
In this section we provide the coding scheme and the upper bound for
the case when there is no public discussion.
\subsection{Proof of Theorem~\ref{thm:lbNoDisc}}
A sequence of length $n$ code is described as follows.

\subsubsection{Codebook Generation}
\begin{itemize}
\item Generate a total of  $2^{n(I(\rvu;\rvy_e) -
2\eps_n)}$ sequences. Each sequence is sampled i.i.d.\ from a
distribution $p_\rvu(\cdot)$.

\item Select a rate $R = I(\rvu;\rvyr)- I(\rvu;\rvye)-\eps_n$ and
randomly partition the set sequences in the previous step into
$2^{nR}$ bins so that there are $2^{n(I(\rvu;\rvyr)-\eps_n)}$
sequences in each bin.
\end{itemize}

\subsubsection{Encoding}
\begin{itemize}
\item Given a state sequence $\rvs^n$ the encoder selects a sequence
$\rvu^n$ randomly from the list of all possible sequences that are
jointly typical with $\rvs^n$.
\item At time $i=1,2,\ldots, n$ the encoder transmits symbol
$\rvx_i$ generated by sampling the distribution
$p_{\rvx|\rvu,\rvs}(\cdot| u_i, s_i)$.
\end{itemize}

\subsubsection{Secret-key generation}
\begin{itemize}
\item The decoder upon observing $\rvyr^n$  finds a sequence
$\rvu^n$ jointly typical with $\rvyr^n$.
\item Both encoder and the decoder declare the bin-index of $\rvu^n$
to be the secret-key.
\end{itemize}

\subsubsection{Secrecy Analysis}
We need to show that for the proposed encoder and decoder, the
equivocation at the eavesdropper satisfies
\begin{equation}
\label{eq:equivCondn}\frac{1}{n}H(\rvk|\rvye^n)= I(\rvu;\rvyr)-
I(\rvu;\rvye) + o_n(1),
\end{equation}where $o_n(1)$ is a term that goes to zero as $n\rightarrow
\infty$.

Accordingly note that \begin{align*} \frac{1}{n}H(\rvk|\rvye^n)&=
\frac{1}{n}H(\rvk,\rvu^n|\rvye^n) - \frac{1}{n}H(\rvu^n|\rvye^n,
\rvk)\\
&= \frac{1}{n}H(\rvu^n|\rvye^n) - \frac{1}{n}H(\rvu^n|\rvye^n,
\rvk)\\
&= \frac{1}{n}H(\rvu^n|\rvye^n) - \eps_n\\
\end{align*}
where the last step follows from the fact that there are at-most
$2^{n(I(\rvu;\rvye)-o_n(1))}$ sequences in each bin and hence the
eavesdropper can decode the codeword $\rvu^n$ given the key $\rvk$.
It remains to lower-bound the first conditional entropy term.
\begin{align}
&\frac{1}{n}H(\rvu^n|\rvye^n) = \frac{1}{n}H(\rvu^n) +
\frac{1}{n}H(\rvye^n|\rvu^n)- \frac{1}{n}H(\rvye^n)\\
&= \frac{1}{n}H(\rvu^n) + \frac{1}{n}H(\rvye^n|\rvu^n,\rvs^n)-
\frac{1}{n}H(\rvye^n) + \frac{1}{n}I(\rvs^n;\rvye^n|\rvu^n)
\label{eq:condEntLB}
\end{align}
We now appropriately bound each term in~\eqref{eq:condEntLB}. First
note that since the sequence $\rvu^n$ is uniformly distributed among
the set of all possible codeword sequences, it follows that
\begin{align}
\frac{1}{n}H(\rvu^n)&= \frac{1}{n}\log_2 |\cC| - \eps_n \notag\\
&= I(\rvu;\rvyr) - \eps_n \label{eq:EquivTerm1}
\end{align}

Next, given $(\rvu^n,\rvs^n)$, as verified below, the channel to the
eavesdropper is memoryless:
\begin{align*}
&p_{\rvye^n|\rvu^n,\rvs^n}(y_e^n|u^n,s^n) \\&= \sum_{x^n \in \cX^n}
p_{\rvye^n|\rvu^n,\rvs^n,\rvx^n}(y_e^n|u^n,s^n,x^n)p(\rvx^n|\rvu^n,\rvs^n)(x^n|u^n,s^n)\\
&= \sum_{x^n \in \cX^n} \prod_{i=1}^n
p_{\rvy_{e}|\rvu,\rvs,\rvx}(y_{e,i}|u_i,s_i,x_i)p_{\rvx|\rvu,\rvs}(x_i|u_i,s_i)\\
&= \prod_{i=1}^n
\sum_{x_i \in \cX}p_{\rvy_{e}|\rvu,\rvs,\rvx}(y_{e,i}|u_i,s_i,x_i)p_{\rvx|\rvu,\rvs}(x_i|u_i,s_i)\\
&= \prod_{i=1}^n p_{\rvy_{e}|\rvu,\rvs}(y_{e,i}|u_i,s_i)\\
\end{align*}
The second step above follows from the fact that the channel is
memoryless and the symbol $\rvx_i$ at time $i$ is generated as a
function of $(\rvu_i,\rvs_i)$. Hence we have that
\begin{align}
&\frac{1}{n}H(\rvye^n|\rvs^n,\rvu^n) = \sum_{i=1}^n
H(\rvy_{e,i}|\rvs^n,\rvu^n,\rvy_{e,1}^{i-1})\\
&= \sum_{i=1}^n H(\rvy_{e,i}|\rvs_i,\rvu_i)\label{eq:EquivTerm2}
\end{align}

Furthermore note that
\begin{equation}
\frac{1}{n}H(\rvye^n) \le \sum_{i=1}^n H(\rvy_{ei}).
\label{eq:YeBound}\end{equation}

Finally, in order to lower bound the term $I(\rvs^n;\rvye^n|\rvu^n)$
we let $J$ to be a random variable which equals 1 if
$(\rvs^n,\rvu^n)$ are jointly typical. Note that $\Pr(J=1) = 1 -
o_n(1)$.
\begin{align}
&\frac{1}{n}I(\rvs^n;\rvye^n|\rvu^n) = \frac{1}{n}H(\rvs^n|\rvu^n)
- \frac{1}{n}H(\rvs^m|\rvu^n,\rvye^n) \notag\\
&\ge \frac{1}{n}H(\rvs^n|\rvu^n,J=1)\Pr(J=1)
- \frac{1}{n}H(\rvs^n|\rvu^n,\rvye^n)\notag\\
&\ge \frac{1}{n}H(\rvs^n|\rvu^n,J=1) -
\frac{1}{n}H(\rvs^n|\rvu^n,\rvye^n) - o_n(1) \notag\\
&\ge H(\rvs|\rvu) - \frac{1}{n}H(\rvs^n|\rvu^n,\rvye^n) - o_n(1)
\label{eq:NoMem}\\
&\ge H(\rvs|\rvu) -
\frac{1}{n}\sum_{i=1}^nH(\rvs_i|\rvu_i,\rvy_{e,i}) -
o_n(1)\label{eq:mutInfBound}\end{align} where~\eqref{eq:NoMem}
follows from the fact that $\rvs^n$ is an i.i.d.\ sequence and hence
conditioned on the fact that $(\rvs^n,\rvu^n)$ is a pair of typical
sequence there are $2^{n H(\rvs|\rvu)- no_n(1)}$ possible sequences
$\rvs^n$.

Substituting~\eqref{eq:EquivTerm1},~\eqref{eq:EquivTerm2},~\eqref{eq:YeBound}
and~\eqref{eq:mutInfBound} in the lower bound~\eqref{eq:condEntLB}
and using the fact that as $n \rightarrow \infty$, the summation
converges to the mean values,
\begin{align*}
&\frac{1}{n}H(\rvk|\rvye^n) \\
 &= I(\rvu;\rvyr)+ H(\rvye|\rvu, \rvs)- H(\rvye)+ H(\rvs|\rvu)-
 H(\rvs|\rvu,\rvye)- o_n(1)\\
 &= I(\rvu;\rvyr)-I(\rvye;\rvs|\rvu) - I(\rvye;\rvu)
 +I(\rvye;\rvs|\rvu)-o_n(1)\\
 &= I(\rvu;\rvyr) - I(\rvye;\rvu) - o_n(1)
\end{align*}as required.

\subsection{Proof of Theorem~\ref{thm:ubNoDisc}}
A sequence of length-$n$ code satisfies:
\begin{align}
&\frac{1}{n}H(\rvk|\rvyr^n) \le \eps_n~\label{eq:fano}\\
&\frac{1}{n}H(\rvk|\rvye^n) \ge \frac{1}{n}H(\rvk)-
\eps_n~\label{eq:secrecy}
\end{align}
where~\eqref{eq:fano} follows from the Fano's Lemma since the
receiver is able to recover the secret-key $\rvk$ given $\rvyr^n$
and~\eqref{eq:secrecy} is a consequence of the secrecy constraint.
Furthermore, note that $\rvk \rightarrow (\rvx^n,\rvs^n) \rightarrow
(\rvyr^n,\rvye^n)$ holds as the encoder generates the secret key
$\rvk$. Thus we can bound the rate $R = \frac{1}{n}H(\rvk)$ as
below:
\begin{align*}
nR &\le I(\rvk;\rvyr^n|\rvye^n) + 2n\eps_n\\
&\le I(\rvk,\rvs^n,\rvx^n;\rvyr^n|\rvye^n) + 2n\eps_n\\
&= I(\rvs^n,\rvx^n;\rvyr^n|\rvye^n) + 2n\eps_n\\
&= \sum_{i=1}^n I(\rvs_i,\rvx_i;\rvy_{r,i}|\rvy_{e,i}) + 2n\eps_n\\
&\le nI(\rvx,\rvs;\rvyr|\rvye) + 2n\eps_n
\end{align*}
where the last step follows from the concavity of the conditional
entropy term $I(\rvx,\rvs;\rvyr|\rvye)$ in the input distribution
$p_{\rvx,\rvs}$ (see e.g.,~\cite{khistiTchamWornell:07}).

Finally since the secret-key capacity only depends on the marginal
distribution of the channel and not on the joint distribution we can
minimize over all joint distributions with fixed marginal
distributions.

\section{With Public Discussion}
In this section we provide the proofs of the coding theorem and the
converse for the case when there is a public discussion channel
allowed.

\subsection{Proof of Theorem~\ref{thm:lbDisc}}
Our coding scheme is closely related to the coding theorem for the
\emph{channel  model} in~\cite{ahlswedeCsiszar:93,maurer:93} and
emulates the generation of correlated source sequences. It consists
of the following steps:
\begin{itemize}

\item Fix a distribution $p_{\rvx|\rvs}$. This induces a
joint distribution $p_{\rvx,\rvyr,\rvye,\rvs}$. Let $R =
I(\rvyr;\rvx,\rvs)- I(\rvyr;\rvye) - \eps_n$

\item Partition the set of all typical sequences $\rvyr^n$ into
$2^{n(H(\rvyr|\rvx,\rvs)-o_n(1))}$ bins. Furthermore partition the
collection of $2^{n(I(\rvyr;\rvx,\rvs))}$ sequences in each bin into
further $2^{nR}$ sequences so that there are
$2^{n(I(\rvyr;\rvye)-\eps_n)}$ sequences in each sub-bin.

\item Given symbol $\rvs_i$ at time $i$, sample a symbol $\rvx_i$ from
the conditional distribution $p_{\rvx|\rvs}(\cdot)$ and transmit it
over the channel.

\item The receiver upon observing $\rvyr^n$ transmits the bin index
of this sequence over the channel. Using the bin index and the
knowledge of $(\rvx^n,\rvs^n)$ the sender reproduces $\rvyr^n$.

\item Both the sender and the receiver declare the sub-bin index of
$\rvyr^n$ as the secret-key.
\end{itemize}

Following the secrecy analysis
in~\cite{ahlswedeCsiszar:93,maurer:93} it can be shown that this
construction satisfies the secrecy constraint~\eqref{eq:secInterac}
and furthermore attains a rate of
$$R = I(\rvyr;\rvx,\rvs)-I(\rvyr;\rvye) +  o_n(1).$$

\subsection{Proof of Theorem~\ref{thm:ubDisc}}
We now establish a corresponding upper bound on the secret-key
capacity.

First, using the fact that the receiver is able to recover the
secret-key and the eavesdropper is subjected to a secrecy
constraint~\eqref{eq:secInterac}, we have that
\begin{align}
&\frac{1}{n}H(\rvk|\rvyr^n,\rvv,\Phi^k) \le \eps_n\\
&\frac{1}{n}I(\rvk; \rvye^n,\Psi^k,\Phi^k) \le \eps_n\\
\end{align}
Using the above relations and the fact that $R =
\frac{1}{n}H(\rvk)$, we note that \begin{align} &nR \le H(\rvk) \notag\\
&\le I(\rvk;\rvyr^n,\rvv,\Phi^k)-I(\rvk;\rvye^n,\Phi^k,\Psi^k) +
2n\eps_n \notag\\
&\le I(\rvk;\rvyr^n,\rvv,\rvye^n,\Phi^k,\Psi^k) - I(\rvk;\rvye^n,\Phi^k,\Psi^k) + 2n\eps_n \notag\\
&\le I(\rvk;\rvyr^n,\rvv|\rvye^n,\Phi^k,\Psi^k) + 2n\eps_n \notag\\
&\le I(\rvu,\rvs^n;\rvyr^n,\rvv|\rvye^n,\Phi^k,\Psi^k) + 2n\eps_n \label{eq:kgenSender} \\
&= I(\rvu,\rvs^n;\rvyr^n,\rvv,\rvye^n,\Phi^k,\Psi^k) - I(\rvu,\rvs^n;\rvye^n,\Phi^k,\Psi^k) + 2n\eps_n \notag\\
&= I(\rvu,\rvs^n;\rvv,\Phi^{i_1-1},\Psi^{i_1-1})\notag\\
&\quad+
I(\rvu,\rvs^n;\rvyr^n,\rvye^n, \Phi_{i_1+1}^k, \Psi_{i_1+1}^k|
\Phi^{i_1-1},\Psi^{i_1-1},\rvv) \notag
\\&\quad - I(\rvu,\rvs^n;\Phi^{i_1-1},\Psi^{i_1-1}) \notag\\
&\quad - I(\rvu,\rvs^n;\rvye^n,\Phi_{i_1+1}^k,
\Psi_{i_1+1}^k|\Psi^{i_1-1},\Phi^{i_1-1})\notag\\
&= I(\rvu,\rvs^n;\rvv|\Phi^{i_1-1},\Psi^{i_1-1}) +
\sum_{j=1}^n{F_{r,j}+G_{r,j}} - \sum_{j=1}^n{F_{e,j}+G_{e,j}}
\label{eq:decomp}
\end{align}where we have introduced
\begin{align}
&F_{r,j} = I(\rvu,\rvs^n;\rvy_{r,j},\rvy_{e,j}|\rvv,\phi^{i_j-1},
\psi^{i_j-1}, \rvyr^{j-1},\rvye^{j-1})\label{eq:Freq}\\
&G_{r,j} = I(\rvu,\rvs^n;
\psi_{i_j+1}^{i_{j+1}-1},\phi_{i_j+1}^{i_{j+1}-1}|\rvv,\phi^{i_j-1},\psi^{i_j-1},\rvyr^j,\rvye^j) \label{eq:Greq}\\
&F_{e,j} = I(\rvu,\rvs^n;
\rvy_{e,j}|\phi^{i_j-1},\psi^{i_j-1},\rvv)\label{eq:Feeq}\\
&G_{e,j} = I(\rvu,\rvs^n;
\phi_{i_j+1}^{i_{j+1}-1},\psi_{i_j+1}^{i_{j+1}-1}|\phi^{i_j-1},\psi^{i_j-1},\rvye^j)
\label{eq:Geeq}\end{align}

To complete the proof, it suffices to show that the following
relations in~\eqref{eq:decomp} hold
\begin{align}
&I(\rvu,\rvs^n;\rvv|\Phi^{i_1-1},\Psi^{i_1-1}) = 0 \label{eq:term1}\\
&F_{r,j}-F_{e,j} \le I(\rvx_j,\rvs_j;\rvy_{r,j}|\rvy_{e,j})\label{eq:term2}\\
&G_{r,j}-G_{e,j} \le  0\label{eq:term3}
\end{align}
To establish~\eqref{eq:term1} note that for $0\le k \le i_1-1$ we
have that $\Phi_k = \Phi_k(\rvu,\rvs^n,\Psi^{k-1})$ and likewise
$\Psi_k = \Psi_k(\rvv,\Phi^{k-1})$. Using which
\begin{align*}
& I(\rvu,\rvs^n;\rvv|\Phi^{i_1-1},\Psi^{i_1-1}) \\
&\le I(\rvu\Phi_{i_1-1},\rvs^n;\rvv,\Psi_{i_1-1}|\Phi^{i_1-2},\Psi^{i_1-2}) \\
&= I(\rvu,\rvs^n;\rvv,\Phi^{i_1-2},\Psi^{i_1-2}) \\
\end{align*}
Continuing this process we have that
$$I(\rvu,\rvs^n;\rvv|\Phi^{i_1-1},\Psi^{i_1-1}) \le I(\rvu,\rvs^n;\rvv)=0,$$
where the last relation follows from the fact that $\rvv$ is
independent of $(\rvu,\rvs^n)$.

In order to establish~\eqref{eq:term2}, we use~\eqref{eq:Freq}
and~\eqref{eq:Feeq} to get,
\begin{align}
&F_{r,j}-F_{e,j} \notag\\
&=I(\rvu,\rvs^n;\rvy_{r,j},\rvy_{e,j}|\rvv,\phi^{i_j-1},\psi^{i_j-1},\rvye^{j-1},\rvyr^{j-1})\notag\\
&\quad - I(\rvu,\rvs^n;
\rvy_{e,j}|\phi^{i_j-1},\psi^{i_j-1},\rvye^{j-1})\notag\\
&=
H(\rvy_{r,j},\rvy_{e,j}|\rvv,\phi^{i_j-1},\psi^{i_j-1},\rvye^{j-1},\rvyr^{j-1})
\notag\\ &\quad-
H(\rvy_{r,j},\rvy_{e,j}|\rvv,\phi^{i_j-1},\psi^{i_j-1},\rvye^{j-1},\rvyr^{j-1},\rvu,\rvs^n)
\notag\\
&\quad - H(\rvy_{e,j}|\phi^{i_j-1},\psi^{i_j-1},\rvye^{j-1}) \notag\\
&\quad + H(\rvy_{e,j}|\phi^{i_j-1},\psi^{i_j-1},\rvye^{j-1},\rvu,\rvs^n)\notag \\
&=
H(\rvy_{r,j},\rvy_{e,j}|\rvv,\phi^{i_j-1},\psi^{i_j-1},\rvye^{j-1},\rvyr^{j-1})
\notag\\ &\quad-
H(\rvy_{r,j},\rvy_{e,j}|\rvv,\phi^{i_j-1},\psi^{i_j-1},\rvye^{j-1},\rvyr^{j-1},\rvu,\rvs^n,\rvx_j)
\notag\\
&\quad - H(\rvy_{e,j}|\phi^{i_j-1},\psi^{i_j-1},\rvye^{j-1}) \notag\\
&\quad + H(\rvy_{e,j}|\phi^{i_j-1},\psi^{i_j-1},\rvye^{j-1},\rvu,\rvs^n,\rvx_j)\label{eq:xcondn} \\
&\le H(\rvy_{r,j},\rvy_{e,j}|\phi^{i_j-1},\psi^{i_j-1},\rvye^{j-1})
- H(\rvy_{r,j},\rvy_{e,j}|\rvx_j,\rvs_j)
\notag\\
&\quad - H(\rvy_{e,j}|\phi^{i_j-1},\psi^{i_j-1},\rvye^{j-1}) + H(\rvy_{e,j}|\rvx_j,\rvs_j) \label{eq:channelMem}\\
&\le H(\rvy_{r,j}|\rvy_{e,j}) -
H(\rvy_{r,j}|\rvy_{e,j},\rvx_j,\rvs_j) =
I(\rvs_j,\rvx_j;\rvy_{r,j}|\rvy_{e,j})
\end{align}
In the above steps~\eqref{eq:xcondn} follows from the fact that
$\rvx_j = X_j(\rvu,\rvs^n,\psi^{i_j-1})$ and hence we can condition
of $\rvx_j$ i the second and fourth terms. Furthermore since the
channel is memoryless $$(\rvy_{r,j},\rvy_{e,j})\rightarrow
(\rvx_j,\rvs_j) \rightarrow
(\rvv,\rvu,\rvs_{j+1}^n,\rvs_{1}^{j-1},\phi^{i_j-1},\psi^{i_j-1},\rvye^{j-1},\rvyr^{j-1})$$
holds.

It remains to establish~\eqref{eq:term3}. Using~\eqref{eq:Greq}
and~\eqref{eq:Geeq} we note that
\begin{align*}
&G_{r,j}-G_{e,j}  \\
&=I(\rvu,\rvs^n;
\psi_{i_j+1}^{i_{j+1}-1},\phi_{i_j+1}^{i_{j+1}-1}|\rvv,\phi^{i_j-1},\psi^{i_j-1},\rvyr^j,\rvye^j)
 \\&\quad - I(\rvu,\rvs^n;
\phi_{i_j+1}^{i_{j+1}-1},\psi_{i_j+1}^{i_{j+1}-1}|\phi^{i_j-1},\psi^{i_j-1},\rvye^j)\\
&= H(\rvu,\rvs^n|\rvv,\phi^{i_j-1},\psi^{i_j-1},\rvyr^j,\rvye^j)\\
 &\quad - H(\rvu,\rvs^n|\rvv,\phi^{i_{j+1}-1},\psi^{i_{j+1}-1},\rvyr^j,\rvye^j)\quad\\
 &\quad - H(\rvu,\rvs^n|\phi^{i_j-1},\psi^{i_j-1},\rvye^j) +
 H(\rvu,\rvs^n|\phi^{i_{j+1}-1},\psi^{i_{j+1}-1},\rvye^j)\\
&= I(\rvu,\rvs^n;\rvv,\rvyr^j|\phi^{i_{j+1}-1},\psi^{i_{j+1}-1},\rvye^j) - I(\rvu,\rvs^n;\rvv,\rvyr^j|\rvye^j,\phi^{i_j-1},\psi^{i_j-1})\\
\end{align*}
Since $\phi_{i_{j+1}-1}=
\Phi_{i_{j+1}-1}(\rvu,\rvs^n,\psi^{i_{j+1}}-2)$ and
$\psi_{i_{j+1}-1}= \Psi_{i_{j+1}-1}(\rvv,\rvyr^j,\phi^{i_{j+1}}-2)$
we have that \begin{align*} &
I(\rvu,\rvs^n;\rvv,\rvyr^j|\phi^{i_{j+1}-1},\psi^{i_{j+1}-1},\rvye^j)\\
&\le
I(\rvu,\rvs^n,\phi_{i_{j+1}-1};\rvv,\rvyr^j,\psi_{i_{j+1}-1}|\phi^{i_{j+1}-2},\psi^{i_{j+1}-2},\rvye^j)\\
&=I(\rvu,\rvs^n;\rvv,\rvyr^j|\phi^{i_{j+1}-2},\psi^{i_{j+1}-2},\rvye^j)
\end{align*} and continuing this process we have that
$$I(\rvu,\rvs^n;\rvv,\rvyr^j|\phi^{i_{j+1}-1},\psi^{i_{j+1}-1},\rvye^j)
\le I(\rvu,\rvs^n;\rvv,\rvyr^j|\rvye^j,\phi^{i_j-1},\psi^{i_j-1})$$
as required.
\section{Gaussian Case}
In this section we develop the corresponding results for the
Gaussian case.
\subsection{Proof of Prop.~\ref{prop:lbGaussNoDisc}}
The lower bound expression follows from Theorem~\ref{thm:lbNoDisc}
by choosing $\rvx \sim \cN(0,P)$ to be a Gaussian random variable
independent of $\rvs$ and by choosing $\rvu = \rvx + \alpha \rvs$.
In this case,
\begin{align*}
R &= I(\rvu;\rvyr) - I(\rvu;\rvye) \\
&= h(\rvu|\rvye) - h(\rvu|\rvyr)\end{align*}
Further evaluating each of the terms above with $\rvu = \rvx + \al \rvs$, note that
\begin{align*}&h(\rvu|\rvye)=\\& \frac{1}{2}\log\left(P + \al^2 Q + 2\al\rho \sqrt{PQ}-
\frac{(P+\al Q + (1+\al)\rho\sqrt{PQ})^2}{P+Q+1+\Delta + 2\rho\sqrt{PQ}}\right)\end{align*}and\begin{align*}
&h(\rvu|\rvyr)=\\&\quad \frac{1}{2}\log\left(P + \al^2 Q + 2\al\rho\sqrt{PQ}-
\frac{(P+\al Q + \rho(1+\al)\sqrt{PQ})^2}{P+Q+1+2\sqrt{PQ}}\right).\end{align*}
This yields that
\begin{multline}
R= \frac{1}{2}\log\left(1 + \frac{\Delta}{1
+\frac{PQ(\al-1)^2(1-\rho^2)}{P+ \al^2Q +2\rho\al\sqrt{PQ}}}\right)\\
\quad +\frac{1}{2}\log\left(\frac{P+Q+1 + 2\rho\sqrt{PQ}}{P+Q+1+\Delta+2\rho\sqrt{PQ}}\right).
\end{multline}
Note that the first term in the expression above is maximized when $\al =1$. As we show below, this choice is indeed feasible when $P\ge 1$.  
In particular the constraint~\eqref{eq:lowerBoundNoDiscCons} requires
that
\begin{align*}
&h(\rvu|\rvs) \ge h(\rvu|\rvyr)\\
& \Rightarrow \frac{1}{2}\log P(1-\rho^2) \ge \\
&\frac{1}{2}\log\left(\frac{PQ(\al-1)^2(1-\rho^2) + (P+
\al^2  Q) + 2\rho\al\sqrt{PQ})}{P+Q+1+2\rho\sqrt{PQ}}\right).
\end{align*}
Substituting $\al = 1$ above we have that
\begin{align}
P(1-\rho^2) &\ge 1-\frac{1}{P+Q+1 + 2\rho\sqrt{PQ}} \ge 1-\frac{1}{P+Q+1}
\end{align}
as required.
\subsection{Proof of Prop.~\ref{prop:ubGaussNoDisc}}

We evaluate the upper bound in Theorem~\ref{thm:ubNoDisc} for the
choice $\rvze = \rvzr + \rvz_\delta$, where $\rvz_\delta \sim
\cN(0,\Delta)$ is independent of $\rvzr$.

\begin{align*}
&I(\rvs,\rvx;\rvyr|\rvye) = h(\rvyr|\rvye)-
h(\rvyr|\rvye,\rvx,\rvs)\\
&= h(\rvyr|\rvye)-
h(\rvzr|\rvze\\
&\le \frac{1}{2}\log\left(P+Q+1 + 2\sqrt{PQ}-
\frac{(P+Q+1 + 2\sqrt{PQ})^2}{P+Q+1+\Delta + 2\sqrt{PQ}}\right)-
\\&\quad-\frac{1}{2}\log\left(1- \frac{1}{1+\Delta}\right)
\end{align*}
where we have used the fact that the conditional entropy
$h(\rvyr|\rvye)$ is maximized by a Gaussian distribution. The above
expression gives~\eqref{eq:ubGaussNoDisc}.

\subsection{Proof of Proposition~\ref{prop:DPCwDC}}
Since the Gaussian model satisfies the condition in
Theorem~\ref{thm:capdisc}, it suffices to evaluate $C=
I(\rvx,\rvs;\rvyr|\rvye)$.
\begin{align}
&I(\rvx,\rvs;\rvyr|\rvye) = h(\rvyr|\rvye)-
h(\rvyr|\rvye,\rvx,\rvs)\\
&= h(\rvyr|\rvye)- h(\rvzr|\rvze)\\
&= \frac{1}{2}\log 2\pi
e\left(P+Q+1+2\sqrt{PQ}-\frac{(P+Q +2\sqrt{PQ})^2}{P+Q+1+2\sqrt{PQ}+\Delta}\right)
 -\frac{1}{2}\log 2\pi
e
\end{align}
which upon simplifying yields the desired expression.

\bibliographystyle{IEEEtran}

\end{document}

%% file: gww_chars.tex
\newcommand{\cC}{{\mathcal{C}}}

\newcommand{\cI}{{\mathcal{I}}}

\newcommand{\cN}{{\mathcal{N}}}

\newcommand{\cP}{{\mathcal{P}}}

\newcommand{\cX}{{\mathcal{X}}}

\newcommand{\al}{\alpha}

\newcommand{\eps}{\varepsilon}



%% file: gww_defs.tex
\usepackage{verbatim} 
 \usepackage{amsxtra}  
\newcounter{actr}
{\begin{list}{(\alph{actr})}{\usecounter{actr}}}{\end{list}}

\newcounter{ictr}
{\begin{list}{(\roman{ictr})}{\usecounter{ictr}}}{\end{list}}

\newtheorem{thm}{Theorem}

\newtheorem{prop}{Proposition}

\newenvironment{new-proof}[1]
{{\em Proof  #1: }}%
{ \noindent\qed }
\newcommand{\qed}{\rule[0.1ex]{1.4ex}{1.6ex}}

\newcommand{\mrm}{\mathrm}

%% file: gww_rvs.tex
\DeclareMathAlphabet{\mathbsf}{OT1}{cmss}{bx}{n}
\DeclareMathAlphabet{\mathssf}{OT1}{cmss}{m}{sl}

\DeclareSymbolFont{bsfletters}{OT1}{cmss}{bx}{n}  
\DeclareSymbolFont{ssfletters}{OT1}{cmss}{m}{n}
\DeclareMathSymbol{\bsfGamma}{0}{bsfletters}{'000}
\DeclareMathSymbol{\ssfGamma}{0}{ssfletters}{'000}
\DeclareMathSymbol{\bsfDelta}{0}{bsfletters}{'001}
\DeclareMathSymbol{\ssfDelta}{0}{ssfletters}{'001}
\DeclareMathSymbol{\bsfTheta}{0}{bsfletters}{'002}
\DeclareMathSymbol{\ssfTheta}{0}{ssfletters}{'002}
\DeclareMathSymbol{\bsfLambda}{0}{bsfletters}{'003}
\DeclareMathSymbol{\ssfLambda}{0}{ssfletters}{'003}
\DeclareMathSymbol{\bsfXi}{0}{bsfletters}{'004}
\DeclareMathSymbol{\ssfXi}{0}{ssfletters}{'004}
\DeclareMathSymbol{\bsfPi}{0}{bsfletters}{'005}
\DeclareMathSymbol{\ssfPi}{0}{ssfletters}{'005}
\DeclareMathSymbol{\bsfSigma}{0}{bsfletters}{'006}
\DeclareMathSymbol{\ssfSigma}{0}{ssfletters}{'006}
\DeclareMathSymbol{\bsfUpsilon}{0}{bsfletters}{'007}
\DeclareMathSymbol{\ssfUpsilon}{0}{ssfletters}{'007}
\DeclareMathSymbol{\bsfPhi}{0}{bsfletters}{'010}
\DeclareMathSymbol{\ssfPhi}{0}{ssfletters}{'010}
\DeclareMathSymbol{\bsfPsi}{0}{bsfletters}{'011}
\DeclareMathSymbol{\ssfPsi}{0}{ssfletters}{'011}
\DeclareMathSymbol{\bsfOmega}{0}{bsfletters}{'012}
\DeclareMathSymbol{\ssfOmega}{0}{ssfletters}{'012}

\newcommand{\rvk}{{\mathssf{k}}}	

\newcommand{\rvl}{{\mathssf{l}}}	

\newcommand{\rvs}{{\mathssf{s}}}	

\newcommand{\rvu}{{\mathssf{u}}}	

\newcommand{\rvv}{{\mathssf{v}}}	

\newcommand{\rvx}{{\mathssf{x}}}	

\newcommand{\rvy}{{\mathssf{y}}}	
\newcommand{\rvyr}{{\mathssf{y}}_\mrm{r}}	
\newcommand{\rvye}{{\mathssf{y}}_\mrm{e}}	

\newcommand{\rvz}{{\mathssf{z}}}	%
\newcommand{\rvzr}{{\mathssf{z}}_\mathrm{r}}	
\newcommand{\rvze}{{\mathssf{z}}_\mathrm{e}}	




%% file: dpcArxiv.bbl
\begin{thebibliography}{10}
\providecommand{\url}[1]{#1}
\csname url@samestyle\endcsname
\providecommand{\newblock}{\relax}
\providecommand{\bibinfo}[2]{#2}
\providecommand{\BIBentrySTDinterwordspacing}{\spaceskip=0pt\relax}
\providecommand{\BIBentryALTinterwordstretchfactor}{4}
\providecommand{\BIBentryALTinterwordspacing}{\spaceskip=\fontdimen2\font plus
\BIBentryALTinterwordstretchfactor\fontdimen3\font minus
  \fontdimen4\font\relax}
\providecommand{\BIBforeignlanguage}[2]{{%
\expandafter\ifx\csname l@#1\endcsname\relax
\typeout{** WARNING: IEEEtran.bst: No hyphenation pattern has been}%
\typeout{** loaded for the language `#1'. Using the pattern for}%
\typeout{** the default language instead.}%
\else
\language=\csname l@#1\endcsname
\fi
#2}}
\providecommand{\BIBdecl}{\relax}
\BIBdecl

\bibitem{wyner:75Wiretap}
A.~D. Wyner, ``The wiretap channel,'' \emph{Bell Syst.\ Tech.\ J.}, vol.~54,
  pp. 1355--87, 1975.

\bibitem{maurer:93}
U.~M. Maurer, ``Secret key agreement by public discussion from common
  information,'' \emph{IEEE Trans.\ Inform.\ Theory}, vol.~39, pp. 733--742,
  Mar. 1993.

\bibitem{ahlswedeCsiszar:93}
R.~Ahlswede and I.~Csisz{\'a}r, ``Common randomness in information theory and
  cryptography -- {P}art {I}: {S}ecret sharing,'' \emph{IEEE Trans.\ Inform.\
  Theory}, vol.~39, pp. 1121--1132, Jul. 1993.

\bibitem{gelfandPinsker:80}
S.~I. Gel'fand and M.~S. Pinsker, ``Coding for channels with random
  parameters,'' \emph{Problems of Control and Information Theory}, vol.~9, pp.
  19--31, 1980.

\bibitem{costa:83}
M.~H. Costa, ``Writing on dirty paper,'' \emph{IEEE Trans.\ Inform.\ Theory},
  vol.~29, pp. 439--441, May 1983.

\bibitem{mitrapant:06}
C.~Mitrapant, H.~Vinck, and Y.~Luo, ``An achievable region for the gaussian
  wiretap channel with side information,'' \emph{IEEE Trans.\ Inform.\ Theory},
  vol.~52, pp. 2181--2190, May 2006.

\bibitem{LiuChen:07}
W.~Liu and B.~Chen, ``Wiretap channel with two-sided state information,'' in
  \emph{Proc.\ 41st Asilomar Conf.\ on Signals, Systems and Comp.}, Nov. 2007.

\bibitem{steinberg:08}
Y.~Steinberg, ``Simultaneous transmission of data and state with common
  knowledge,'' in \emph{Proc.\ Int.\ Symp.\ Inform.\ Theory}, Toronto, Canada,
  Jul. 2008, pp. 935--939.

\bibitem{steinberg:09}
------, ``Coding and common reconstruction,'' \emph{IEEE Trans.\ Inform.\
  Theory}, vol.~55, pp. 4995--5010, Nov. 2009.

\bibitem{khistiDigWornell:09}
A.~Khisti, S.~N. Diggavi, and G.~W. Wornell, ``Secret-key agreement using
  asymmetric in channel state information,'' in \emph{Proc.\ Int.\ Symp.\
  Inform.\ Theory}, 2009.

\bibitem{prabhakaran:09}
\BIBentryALTinterwordspacing
V.~Prabhakaran, K.~Eswaran, and K.~Ramchandran, ``Secrecy via sources and
  channels,'' \emph{IEEE Trans.\ Inform.\ Theory}, submitted, Nov 2009.
  [Online]. Available:
  \url{http://www.ifp.illinois.edu/~vinodmp/publications/Secrecy09.pdf}
\BIBentrySTDinterwordspacing

\bibitem{chenVinck:06}
Y.~Chen and H.~Vinck, ``Wiretap channel with side information,'' in
  \emph{Proc.\ Int.\ Symp.\ Inform.\ Theory}, Jun. 2006.

\bibitem{khistiTchamWornell:07}
A.~Khisti, A.~Tchamkerten, and G.~W. Wornell, ``Secure {B}roadcasting over
  {F}ading {C}hannels,'' \emph{IEEE Trans.\ Inform.\ Theory, Special Issue on
  Information Theoretic Security}, pp. 2453--2469, 2008.

\end{thebibliography}
